# First results from a second generation galactic axion experiment

C. Hagmann, D. Kinion, W. Stoeffl, K. van Bibber
Lawrence Livermore National Laboratory
7000 East Ave, Livermore, CA 94550

E. Daw, J. McBride, H. Peng, L.J. Rosenberg, H. Xin
Dept of Physics, Massachusetts Institute of Technology
77 Massachusetts Ave, Cambridge, MA 02139

J. Laveigne, P. Sikivie, N.S. Sullivan, D.B. Tanner
Dept of Physics, University of Florida
Gainesville, FL 94720

D. Moltz
Nuclear Science Division, Lawrence Berkeley Laboratory
1 Cyclotron Rd, Berkeley, CA 94720

F. Nezrick, M.S. Turner
Fermi National Accelerator Laboratory
Batavia, IL 60510

N. Golubev, L. Kravchuk
Institute for Nuclear Research of the Russian Academy of Sciences
60th October Anniversary Prospekt 7a, Moscow 117312, Russia

astro-ph/9607022

We report first results from a large-scale search for dark matter axions. The experiment probes axion masses of 1.3-13 $\mu$eV at a sensitivity which is about 50 times higher than previous pilot experiments. We have already scanned part of this mass range at a sensitivity better than required to see at least one generic axion model, the KSVZ axion. Data taking at full sensitivity commenced in February 1996 and scanning the proposed mass range will require three years.

## 1. INTRODUCTION

At present, the hypothetical axion is one of the leading candidates for the dark matter in the universe. The axion is a consequence of the Peccei-Quinn mechansim to solve the strong CP problem in QCD [1,2], and it has properties similar to those of a light neutral pion, but with much weaker couplings to ordinary matter. Cosmological and astrophysical arguments [3] constrain its mass to $10^{-6} < m_a < 10^{-3}$ eV. The relic axion density is roughly inversely proportional to the axion mass and provides critical density for $m_a = O(10^{-6}$ eV). These axions are cold dark matter (CDM) and have never been in thermal equilibrium with other particles. CDM is a major ingredient of theories of structure formation and in particular could be the material that makes up the dark halo of our galaxy.

## 2. TECHNIQUE

Galactic axions are nonrelativistic and can be converted into monochromatic microwave photons ($hf=mc^2(1+O(10^{-6}))$) via the Primakoff effect in a static magnetic field B. This process is resonantly enhanced in a high-Q microwave cavity tuned to the axion mass. The expected signal, normalized to our current experimental parameters, is [4,5]

$$P_{a\rightarrow\gamma} \approx 10^{-21}\,\text{W}\left(\frac{B}{7.7\,\text{T}}\right)^2\left(\frac{V}{200l}\right)\left(\frac{C}{0.65}\right)\left(\frac{Q}{90000}\right)\left(\frac{f}{0.7\,\text{GHz}}\right)\left(\frac{\rho_a}{\rho_{\text{halo}}}\right) \quad (1)$$

where $V$ is the cavity volume, $C$ is a mode dependent form factor (maximal for the fundamental $TM_{010}$ mode), $Q$ is the loaded quality factor, and $\rho_a$ is the local axion density. This power is calculated for the hadronic KSVZ axion [6] and corresponds to a few thousand converted axions per second. The GUT inspired DSFZ axion [7] yields about seven times fewer conversions.

The signature of an axion signal is a narrow peak (of fractional width $\approx 10^{-6}$ due to kinetic energy) above the thermal background composed of cavity blackbody radiation and amplifier noise. The latter is usually expressed as a system noise temperature $T_s=T_c+T_a$, where $T_c$ is the cavity temperature and $T_a$ is the amplifier noise temperature.

Since the axion mass is not known, the cavity must be tunable, so it may be slowly scanned in frequency. For a signal-to-noise ratio s/n and a signal power given by (1), the scan rate is

$$\frac{df}{dt} \approx 25\,\frac{\text{MHz}}{\text{month}}\left(\frac{4}{s/n}\right)^2\left(\frac{6\,\text{K}}{T_s}\right)^2. \quad (2)$$

## 3. DETECTOR

Figure 1 shows the dewar, which contains the magnet and the insert comprising the microwave cavity and amplifier. The magnet has its own helium reservoir and remains cold during changes in the cavity/amplifier configuration. The cylindrical cavity is operated in the $TM_{010}$ mode. It is made of copper plated stainless steel and contains two tuning rods (alumina or copper) which are moved transversely. Using different combinations of rods, the tuning range is (0.65-1.7)$f_0$, where $f_0$ is the resonance frequency of the empty cavity (460 MHz for the first single cavity). A pair of stepper motors on the dewar top drive 1:30000 worm gear reducers mounted on the cavity. The frequency tuning resolution is a few 100 Hz. The resonance frequency (FWHM $\approx$ 10 kHz) is typically stepped by a few kHz.

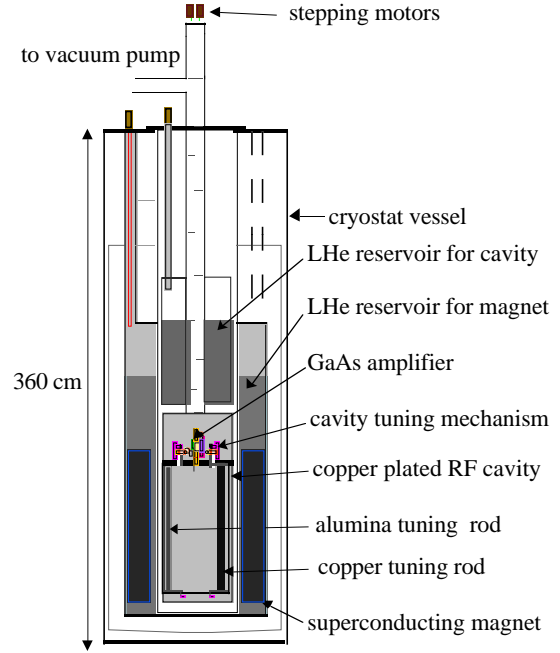

insensitive to the cavity output impedance and avoids the use of a circulator.

Both cavity and amplifier are cooled to 1.4 K with helium gas pumped by a roots blower. A small reservoir of superfluid helium is maintained at the bottom of the can containing the cavity. The liquid level is regulated by a computer controlled valve connected to the main tank. The helium consumption of the cavity is about 7 liters/day.

The magnet is a high inductance (500 H), low current (225 A) NbTi solenoid. The current is provided by a computer controlled power supply with a short term stability of 5ppm. The current runs through a pair of optimized vapor cooled current leads made from copper foil. The dewar has a $LN_2$ jacket and three intermediate temperature layers at 80, 50, and 30 K. The helium usage of the magnet is 50 liters/day.

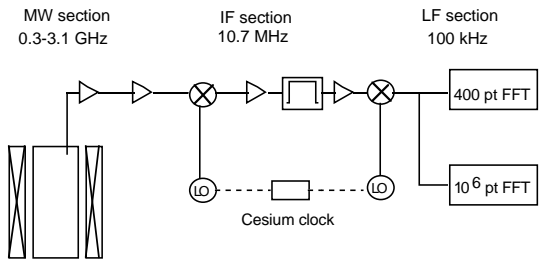

Fig.2 Electronic block diagram.

## 4. SIGNAL PROCESSING

The signal processing chain is shown in Figure 2. It consists of several amplification and mixing stages. The low frequency signal is spectrum analyzed at two different resolutions: a medium resolution ($\Delta f$=100 Hz, $\Delta f/f \approx 10^{-7}$, 10000 averages), and a high resolution channel ($\Delta f$=0.02 Hz, $\Delta f/f \approx 10^{-11}$, no averages). The latter was implemented to look for possible fine structure [8,9] in the galactic axion line. The integration time per tuning rod setting was set to about 50 seconds in order to obtain the scan rate in (2). Each frequency range is scanned three times to minimize the contributions of spurious noise. The FFT spectra are periodically logged to disk for off line analysis and peak search. The experiment is fully computer controlled by a Macintosh computer running Labview. Since 2/96, we have been running with a duty cycle of > 90 %.

## 5. CALIBRATION OF SENSITIVITY

The sensitivity of the axion detector is measured with the variable temperature method. The cavity, which is critically coupled, acts as a 50 Ω termination and emits a noise power $P_c = k_B T B$, where B is the bandwidth and $k_B$ is Boltzmann's constant. The noise power at the output of the amplifier with gain G is $P = k_B B G (T_c + T_a)$. The amplifier temperature $T_a$ can be determined by measuring P while varying the cavity temperature $T_c$, since B, G, and $T_a$ are constant. Figure 3 shows the data from a calibration run at f=700MHz. The system noise temperature is therefore $T_s = T_a + T_c = 4.3$ K + 1.4 K = 5.7 K.

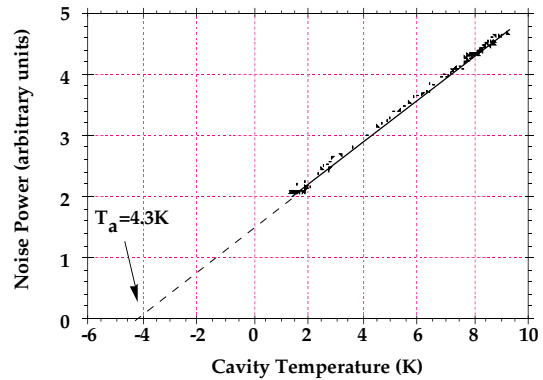

Fig.3 Calibration of amplifier noise temperature by the cavity heating method.

## 6. REACH OF EXPERIMENT

We intend to cover a decade of axion masses (1.3-13 µeV) in about three years. Our sensitivity is sufficient to probe the halo for axions with theoretically predicted couplings. The proposed mass region is shown in Figure 4. Also shown is the already excluded parameter space from previous pilot experiments [10,11]. To date (5/96), we have scanned the region 660-720 MHz (2.7-3.0 µeV), indicated in the plot. Data analysis is still in progress.

For frequencies above 800 MHz (3.3 µeV), an array of 4 or 16 identical cavities will be used. These cavities must be synchronously stepped and their ouput power-combined in phase. The hardware is already under construction and the switchover to the 4-array will occur in early 1997.

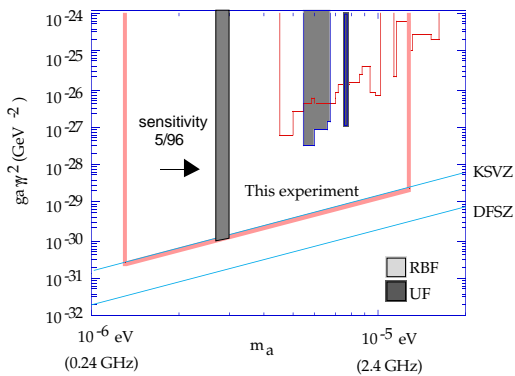

Fig.4 Exclusion plot for galactic axions, assuming a halo density $\rho_a$ = 300 MeV/cm$^3$. The heavy dark line indicates where we have already scanned down to KSVZ sensitivity, but data analysis is still in progress.

## ACKNOWLEDGEMENTS

This work was performed under the auspices of the U.S. Department of Energy under contracts no. W-7405-ENG-48 (LLNL), DE-AC03-76SF00098 (LBL), DE-AC02-76CH03000 (FNAL), DE-FC02-94ER40818 (MIT), DE-FG05-86ER40272 (UF), and FG02-90ER-40560 (U Chicago).

## REFERENCES

1. R. Peccei and H. Quinn, *Phys. Rev. Lett.* **38** (1977) 1440.
2. S. Weinberg, *Phys. Rev. Lett.* **40** (1978) 223; F. Wilczek, *Phys. Rev. Lett.* **40** (1978) 279.
3. For a review see M. Turner, *Phys. Rep.* **197** (1990) 67.
4. P. Sikivie, *Phys. Rev. Lett.* **51** (1983) 1415.
5. L. Krauss et al., *Phys. Rev. Lett.* **55** (1985) 1797.
6. J. Kim, *Phys. Rev. Lett.* **43** (1979) 103; M. Shifman et al., *Nucl. Phys.* **B166** (1980) 493.
7. M. Dine et al., *Phys. Lett.* **104B** (1981) 199; A. Zhitnitskii, *Sov. J. Nucl. Phys.* **31** (1980) 260.
8. P. Sikivie and J. Ipser, *Phys. Lett.* **B291** (1992) 288.
9. P. Sikivie et al., *Phys. Rev. Lett.* **75** (1995) 2911.
10. W. Wuensch et al., *Phys. Rev.* **D40** (1989) 3153.
11. C. Hagmann et al., *Phys. Rev.* **D42** (1990) 1297.